\documentclass{article}%
\usepackage{amsmath}
\usepackage{amsfonts}
\usepackage{amssymb}
\usepackage{graphicx}%
\providecommand{\U}[1]{\protect\rule{.1in}{.1in}}

\begin{document}
\begin{titlepage}
\vspace*{-2cm}
\begin{flushright}
\end{flushright}
\vspace*{1cm}
{\Large
\begin{center}
\bf{Neutrino masses and proton decay in SO(10)}
\end{center}
}
\vspace*{0.5cm}
\begin{center}
M. Abud$^1,^2$, F. Buccella$^2$ and D. Falcone \\
{\textit{$^1$Universit\'a di Napoli Federico II} } \\
{\textit{$^2$INFN, Sezione di Napoli} }
\vspace*{1cm}
\begin{abstract}
\noindent
We consider the constraints on SO(10) unified models
coming from the lower limits on proton lifetime and
on the scale of B$-$L symmetry breaking within the framework of
the seesaw model for neutrino masses.
By upgrading a triangular relationship for the inverse of $\nu_L$ Majorana
masses to the experimental situation with non maximal $\theta_{23}$ and non
vanishing $\theta_{13}$, we get for the sum of $\nu_L$ masses the upper limit
0.16 eV.
\end{abstract}
\end{center}


\end{titlepage}

\section{Introduction}

It is well known that the minimal $SU(5)$ grand unified model proposed by
Georgi and Glashow \cite{GG} has met a number of shortcomings, the three
running coupling constants do not meet at the same point and, more
importantly, the unification scale of the couplings $g_{2}$ and $g_{1}$ of the
unified electroweak theory is too low to comply with the lower limit on proton
lifetime, which scales as to the fourth power of the unification scale. It was
soon realized that consistency with experiment was obtainable in the extended
$SO(10)$ \cite{GFM} GUT model, provided the intermediate symmetry breaking
pattern comprises a $SU(2)_{R}$ group,contributing to the weak hypercharge
according to $Y=T_{3R}+(B-L)/2$.

In fact, above the intermediate scale the first component belongs to a
non-abelian group and the same happens for the second component, if the
intermediate symmetry contains the Pati-Salam $SU(4)$ \cite{PS}, which
contains \ $SU(3)_{c}\times U(1)_{B-L}$. The change of regime in the RGE
provokes the meeting of the gauge constants at a higher scale. This scale is
even higher, if $D$ parity, which would imply equal values at the intermediate
scale for $g_{2L}$ and $g_{2R}$ is broken at the highest scale \cite{MP}. This
fact induced a systematic study of the scalar potential with absolute minima
in the directions able to provide the desired symmetry breaking, where $B-L$
is broken by the VEV along the $SU(5)$ singlet of the 126 representation,
while the breaking of $SO(10)$ at the highest scale is obtained by a VEV of
the 54 for the case in which the intermediate symmetry is $SU(4)\times
SU(2)\times SU(2)\times D$ \cite{BCW}, and along a particular direction of the
210 in the two-dimensional space of the singlets of $SU(3)\times SU(2)\times
SU(2)\times U(1)$ for the three cases where the intermediate symmetry is
$SU(4)\times SU(2)\times SU(2)$ \cite{BCST} or $SU(3)\times SU(2)\times
SU(2)\times U(1)\times D$ \cite{BR} or just $SU(3)\times SU(2)\times
SU(2)\times U(1)$ \cite{ABRS}.

In the cases of intermediate symmetry group with a $SU(4)$ factor, a larger
(negative) contribution from the gauge bosons imply a more rapid evolution for
the coupling constant $g_{3}$, while in the cases where a discrete $D$
symmetry factor is present, the evolution of $g_{2}L$ is importantly affected
by the more copious scalar content, therefore the intermediate symmetry $SU(3)
\times SU(2) \times SU(2) \times U(1)$ provides the case with the larger
proton lifetime \cite{BMRST}. We assume the Extended Survival Hypothesis (ESH)
\cite{ESH}, which limits the contribution of the scalar particles to the RGE
only to the multiplets containing the Higgses with the VEV's responsible for
the spontaneous symmetry breaking at a lower scale, namely the electroweak
Higgs above $M_{Z}$ and the multiplet responsible for the spontaneous breaking
of the intermediate symmetry between the two highest scales. So we
have four different expressions for the two higher scales in terms of
$\sin^{2} \theta_{W}$ and $\alpha/ \alpha_{s}$ at the electroweak scale. It
has been observed that the scale of $SO(10)$ symmetry breaking, which is
related to proton decay, cannot be larger than the expression in terms of the
difference between $\sin^{2} \theta_{W}$ and $\alpha/ \alpha_{s}$ that one
should get without any change in the evolution. We will give an evaluation of
the corrections implied by consider the RGE at the next order.

More recently \cite{BDLM} an extensive analysis has been performed about
$SO(10)$ models with two intermediate symmetry groups between $SO(10)$ and the
gauge group of the standard model $SU(3) \times SU(2) \times U(1)$. A general
property of these $SO(10)$ models with intermediate scales is that, since the
$Q^{2}$ evolution of $\sin^{2} \theta_{W}$ is more soft just below the scale
of $SO(10)$ breaking, there is often a relationship between the scales of
$SO(10)$ and the $B - L$ symmetry breaking, higher the first, lower the
second. Therefore the models with the highest scale for $M_{X}$, and therefore
longer proton lifetime, is the one with the lowest scale for $B - L$ symmetry
breaking, called $M_{B - L}$.

By assuming the seesaw model \cite{seesaw}, which comes out naturally in the
framework of $SO(10)$ unification and accounts for the order of magnitude of
left-handed neutrino masses, one has the relation
\begin{equation}
\det m_{L} \det M_{R} = \det(m_{D})^{2}%
\end{equation}
where $m_{L} $ and $M_{R}$ are the Majorana mass matrices for left and right
handed neutrinos respectively, and $m_{D}$ is the Dirac neutrino mass matrix.
The relation (1), which is crucial for the lower bound that we will get for the scale of 
$B-L$ symmetry breaking, $M_{B-L}$, holds if one neglects type II seesaw. 
Therefore our result do not apply in presence of non-negligible
$SU(2)_L$ triplet VEV's.
We do not know $m_{D}$, but it is reasonably to assume within the $SO(10)$
model that
\begin{equation}
\det m_{D} \det m_{d} = \det m_{l} \det m_{u},
\end{equation}
at least approximately. This implies
\begin{equation}
\det m_{D} = 2 \cdot10^{7}\ \text{MeV}^{3}%
\end{equation}
so the seesaw model gives
\begin{equation}
\det m_{L} \det M_{R} = \det(m_{D})^{2} = 4 \cdot10^{14}\ \text{MeV}^{6}%
\end{equation}
with the upper limit from cosmology \cite{fogli} $\det m_{L} < 8 \cdot
10^{-3}\ \text{eV}^{3}$, and a corresponding lower limit
\begin{equation}
\det M_{R} > \frac{4 \cdot10^{14}\ \text{MeV}^{6}}{8 \cdot10^{-3}%
\ \text{eV}^{3}} = 0.5 \cdot10^{26}\ \text{GeV}^{3}.
\end{equation}
We expect $\det M_{R}$ to be less than the cube of the scale of $B-L$ symmetry
breaking and so for this scale we find the lower limit $3.68 \times10^{8}$ GeV.

In the four models we are considering, one predicts the values of the two
higher scales of spontaneous symmetry breaking, the highest one being related
to proton decay. Then, the lower limit on proton lifetime implies a
corresponding lower limit for that scale, while the second scale would imply a
lower limit on the product of left-handed neutrino masses. From the general
properties of the $b$'s for the RGE one sees that the trend is such that to a
higher scale for the lepto-quarks mediating proton decay corresponds a lower
scale for the spontaneous symmetry breaking of $B-L$ and consequently a higher
lower limit for the product of left-handed neutrino masses. In conclusion
proton decay and cosmological neutrino masses provide two conflicting limits
on the $SO(10)$ models described here.

\section{Symmetry breaking scales}

The values of $\sin^{2} \theta_{W}$ and $\alpha_{s}$ are $0.23116 \pm0.00013$
and $0.1184 \pm0.0007$, respectively \cite{rpp}. In the four models with
intermediate symmetry containing $SU(2)_{R}$ we have at first order:

\noindent-Model with Higgses in the 54 and intermediate symmetry $SU(4)\times
SU(2)\times SU(2)\times D$
\begin{equation}
3/8-\sin^{2}\theta_{W}=\alpha/2\pi\lbrack109/24\ln(M_{B-L}/M_{Z}%
)-49/24\ln(M_{X}/M_{B-L})]
\end{equation}%
\begin{equation}
3/8-\alpha/\alpha_{s}=\alpha/2\pi\lbrack67/8\ln(M_{B-L}/M_{Z})+49/8\ln
(M_{X}/M_{B-L})]
\end{equation}
-Model with Higgses in the 210 and intermediate symmetry $SU(4)\times
SU(2)\times SU(2)$
\begin{equation}
3/8-\sin^{2}\theta_{W}=\alpha/2\pi\lbrack109/24\ln(M_{B-L}/M_{Z}%
)+11/8\ln(M_{X}/M_{B-L})]
\end{equation}%
\begin{equation}
3/8-\alpha/\alpha_{s}=\alpha/2\pi\lbrack67/8\ln(M_{B-L}/M_{Z})+47/8\ln
(M_{X}/M_{B-L})]
\end{equation}
-Model with Higgses in the 210 and intermediate symmetry $SU(3)\times
SU(2)\times SU(2)\times U(1)_{B-L}\times D$
\begin{equation}
3/8-\sin^{2}\theta_{W}=\alpha/2\pi\lbrack109/24\ln(M_{B-L}/M_{Z}%
)+19/8\ln(M_{X}/M_{B-L})]
\end{equation}%
\begin{equation}
3/8-\alpha/\alpha_{s}=\alpha/2\pi\lbrack67/8\ln(M_{B-L}/M_{Z})+55/8\ln
(M_{X}/M_{B-L})]
\end{equation}
-Model with Higgses in the 210 and intermediate symmetry $SU(3)\times
SU(2)\times SU(2)\times U(1)_{B-L}$
\begin{equation}
3/8-\sin^{2}\theta_{W}=\alpha/2\pi\lbrack109/24\ln(M_{B-L}/M_{Z}%
)+29/12\ln(M_{X}/M_{B-L})]
\end{equation}%
\begin{equation}
3/8-\alpha/\alpha_{s}=\alpha/2\pi\lbrack67/8\ln(M_{B-L}/M_{Z})+25/4\ln
(M_{X}/M_{B-L})].
\end{equation}
By these equations we can obtain the intermediate scale (also the scale of the
mass of RH neutrino) and the unification scale, which is the scale of
leptoquark allowing proton decay. Of course, we should consider the two-loop
equations which were reported for the first time by Jones \cite{Jones}.

To get into account of the two loop contributions, we correct the lowest order
coefficients by just multiplying them by the ratios of the second to the first
order values of \cite{AABPRT}, where the values of the $b_{i}$'s have only the
slight difference of considering two Higgs doublets at the electroweak scale,
while here to avoid flavour changing neutral corrents we have only one. Also
the values of $\sin^{2}\theta_{W}$ and $\alpha_{s}$ are slightly changed, with
a relevant impact on the first order values for the scales and a less
important consequence on the ratios between the second order and first order
values for the scales. Also the corresponding error on the scales is
negligible with respect to the uncertainties on the masses of the scalars, as
it is shown in \cite{AABPRT}. So we get for the scales for the four model
considered the following values:

\begin{center}%
\begin{tabular}
[c]{|c|c|c|}\hline
model & $\log(M_{B-L}/$GeV) & $\log(M_{X}/$GeV)\\\hline
I & 13.65 & 14.98\\
II & 11.08 & 16.06\\
III & 10.28 & 15.51\\
IV & ~9.03 & 16.48\\\hline
\end{tabular}

\end{center}

\vspace{0.5cm}
\noindent
to be compared with the lower limit $M_{X}=10^{15.431}GeV$ coming from the
corresponding lower limit $8\times10^{33}$ years on the rate
$\tau(p\rightarrow e^{+}+\pi^{0})$ \cite{pdecay} and the formula
\begin{equation}
\tau(p\rightarrow e^{+}+\pi^{0})=8\times10^{33}[M_{X}/10^{15.431}%
GeV]^{4} ~\text{ys}.
\end{equation}
The first model gives a lifetime more than an order of magnitude shorter than
the lower limit, which also keeping into account the uncertainties on our
evaluation of the scales strongly disfavours it. The third one gives a
lifetime about a factor two larger than the present lower limit, while the
second and the fourth imply lifetimes not accessible for the next decades. It
is important to stress that due to the fast dependance on $M_{X}$, which has
an exponential dependence on the values of $\sin^{2}\theta_{W}$ and
$\alpha_{s}$ at the scale $M_{Z}$ with a relatively large uncertainty on the
value of $\alpha_{s}$, the conclusions may be changed with larger values for
these constants, which would imply longer lifetimes, while smaller values
would have the opposite effect of shorter lifetimes. Also the uncertainties
associated to the masses of the scalars neglected for the evolution in the ESH
limit the sharpness of our conclusions.

For the four models discussed here we plot in Fig.'s 1-4 the values of the
constants $\sin^{2}\theta_{W}$ and $\alpha_{s}$ with their uncertainties and
the allowed zones for them consistent with the lower limits for the scale
$M_{X}$ and $M_{R}$ coming from the lower limit on proton lifetime and the
upper limit on the sum of the left-handed neutrino masses, which within the
see-saw model (see eq(2)) and assuming eq(3) determine the lower limit for
$M_{R}$, respectively\footnote{For graphic reasons the central bar corresponds
to three standard deviations for $\sin^{2}\theta_{W}$ and only one for
$\alpha_{s}$ .}. The fourth model, the one with the highest $M_{X}$, implies
for neutrino masses values near to the present upper limits, while for the
third one both proton decay and neutrino masses are not so far from the
present limits. Finally one may observe that the present values of the
electro-weak couplings are just in the region, where for the model with
$SU(4)\times SU(2)_{L}\times SU(2)_{R}$ intermediate symmetry experimental
signatures are not expected in the near future.

As long as for the large class of models considered by \cite{BDLM} beyond the
obvious limitation to the models with $n_{U}\geq15.431$ the restriction
$n_{1}\geq8.57$ is affecting also the model with the largest $n_{U}$, also
excluding a large part of the values of $n_{1}$ for the two last models
defined in Table IV of \cite{BDLM}. If $B-L$ is spontaneously broken by a vev
of the 16, one should have for the mass of the Majorana mass of the
right-handed states the expression \cite{W}
\begin{equation}
M_{\nu_{R}}=\left(  \frac{\alpha}{\pi}\right)  ^{2}Y_{10}\frac{M_{B-L}^{2}%
}{M_{X}}%
\end{equation}
$Y_{10}$ being the Yukawa coupling to the 10. The above equation implies that
right-handed neutrino masses are several orders of magnitude smaller than the scale
of spontaneous breaking of $B-L$ and, within our hypotheses, a higher lower
limit for that scale. In conclusion one may say that the models with the
largest $M_{X}$ and therefore longest proton lifetime are the ones, when one
expects the largest signals for $m_{ee}$
(the parameter appearing in neutrinoless double beta decay) and $m_{\nu_{e}}$
(the kinematical neutrino mass related to the final part of the 
electron energy spectrum in tritium decay). Our limits on
$M_{B-L}$ depend of course on our assumption for $\det m_{D}$, but on the
other side we expect that $\det M_{R}$ is smaller than $M_{B-L}^{3}$, which
would make the lower limit on $M_{B-L}$ more restrictive.

In a previous paper \cite{ABFO} from the following assumptions:

1) See-saw model for neutrino masses;

2) A value for the highest eigenvalue of the neutrino Dirac mass matrix
$m_{D}$ of the order of the top-quark mass and a form for the matrix $V^{L}$,
which appears in the biunitary transformation which diagonalizes $m$, similar
to $V_{CKM}$;

3) The upper limit for the mass of the heaviest right-handed neutrino given by
the scale of $B - L$ symmetry breaking in nonsupersymmetric $SO(10)$ model,
which, as it is shown in this paper, is around $10^{11}$ GeV,

\noindent
we derived the sum rule for the inverse of the Majorana masses of the
left-handed neutrinos:
\begin{equation}
\sum_{i}\frac{|U_{\tau i}|^{2}}{m_{i}}=0
\end{equation}
and we considered $\theta_{23}$ maximal and a vanishing $\theta_{13}$.
A finite value of $\theta_{13}$ and
eq(16) have been shown to play an important role \cite{BNardi} for the realization of the leptogenesis scenario for baryogenesis \cite{leptogenesis}.

The Daya Bay \cite{dayabay} and Reno \cite{reno} results confirmed what was
deduced by a global analysis of existing data \cite{fogli13}. With the
parameters in Table 1 of \cite{1205.5254}, namely $\Delta m_{s}^{2}%
=7.54\cdot10^{-5}$ eV$^{2}$, $\Delta m_{a}^{2}=2.43\cdot10^{-3}$ eV$^{2}$, and
$\sin^{2}\theta_{12}=0.307$, $\sin^{2}\theta_{13}=0.0241$, and $\sin^{2}%
\theta_{23}=0.386$, $\delta=1.08\pi$ one has:
\begin{equation}
|U_{\tau1}|^{2}=0.196,~|U_{\tau2}|^{2}=0.204,~|U_{\tau3}|^{2}=0.599,
\end{equation}
with important consequences. It implies the normal hierarchy for the masses of
$\nu_{L}$'s and for $|m_{1}|$ the range: $6.3\cdot10^{-3}\leq|m_{1}%
|\leq4.4\cdot10^{-2}~$eV$.$

One gets an upper limit of $\ 1.28 \cdot 10^{-4} ~\text{eV}^{3}$ \ for $|\det m_{L}|$
$\ \ $and $\ $of $0.16$ eV for the sum of the masses of left-handed neutrinos,
even smaller than the most severe bound, $0.2 ~\text{eV}$ in \cite{fogli}. From eq(4)
the smaller upper limit implies the higher lower limit for $M_{B-L}%
=1.46\cdot10^{9}$ GeV.

In Figures 5, 6, 7 we plot for all the models here considered, excepted the
first one (where the the two lines are external to the diagram) the
constraints following by this stronger limit together with the one
corresponding to $M_{X}$ to a higher lower limit on proton lifetime,
$1.2\cdot10^{34}$ years. With these constraints, the third model would be
almost excluded by mentioned limit on proton decay. The fourth model would be almost excluded, on the other side, by the limit on $M_{B-L}$.

By the way the second model, which is the only one fully consistent with these
more severe constraints, corresponds to a very stable minimun of the Higgs
potential \cite{BCST}.
For this model, with the scale of $B-L$ symmetry breaking around
$10^{11}$ GeV, possible $\Delta (B-L) = - 2$ decays as the ones related
to the $d=7$ effective operators described in \cite{bm}, may be the signal
of baryon non-conservation.

In Figures 8, 9 and 10 we plot $|m_{i}|$ and their sum, $|m_{\nu_{e}}|$ and
$|m_{ee}|$, and $\det(m_L)$ in the allowed range for $|m_{1}|$ given by eq (18).

In conclusion, we obtain the following bounds (all values in meV)
\begin{equation}
63 \le \Sigma m_i \le 155,
\end{equation}

\begin{equation}
11 \le m_{\nu_e} \le 45,
\end{equation}

\begin{equation}
8.6 \le m_{ee} \le 44.7.       
\end{equation}

\textbf{Aknowledgements}

We gratefully aknowledge Prof. G. L. Fogli for his constant encouragement,
and Prof. E. Lisi for an inspiring discussion, which is at the origin of this 
work, and for critically reading this manuscript."

\newpage

\begin{center}

\begin{tabular}{cc}
\includegraphics[scale=0.6]{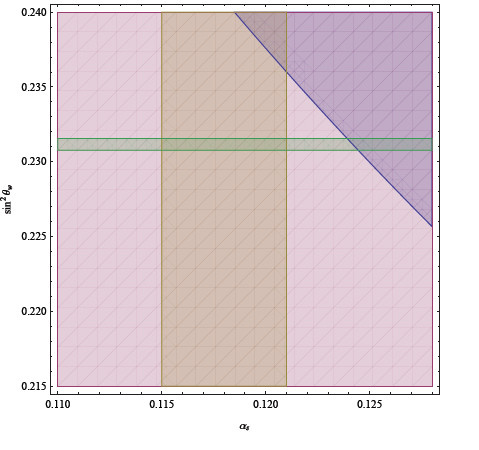} &
\includegraphics[scale=0.6]{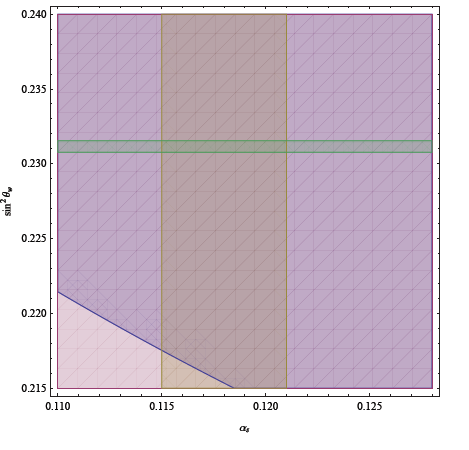} \\ 
Fig. 1 & Fig. 2 \\ 
\end{tabular}

\vspace{1cm}

\begin{tabular}{cc}
\includegraphics[scale=0.6]{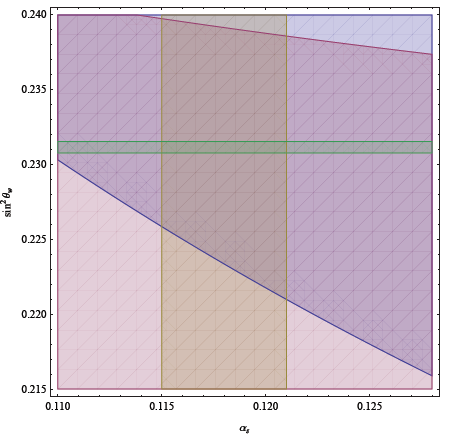} &
\includegraphics[scale=0.6]{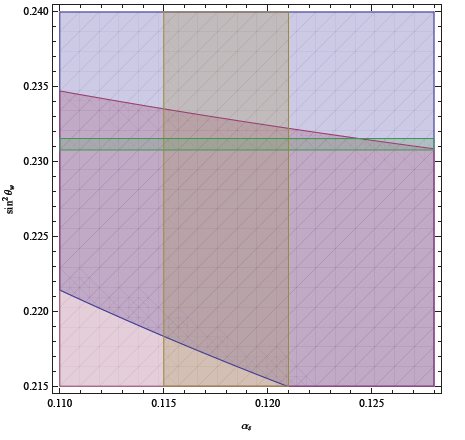} \\ 
Fig. 3 & Fig. 4 \\ 
\end{tabular}

\end{center}

\vspace{1cm}
\noindent
The allowed zones in the ($\alpha_s, \sin^2 \theta_w$)-plane for models I-IV are the ones in the region between the two lines. The blue line is due to the limit on $M_X$, while the red line to the limit on $M_R$.

\newpage

\begin{center}

\begin{tabular}{c}
\hspace{7.5cm} \includegraphics[scale=0.6]{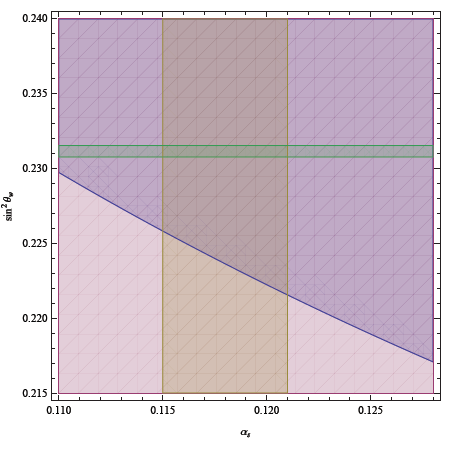} \\ 
\hspace{7.5cm} Fig. 5 \\ 
\end{tabular}

\vspace{1cm}

\begin{tabular}{cc}
\includegraphics[scale=0.6]{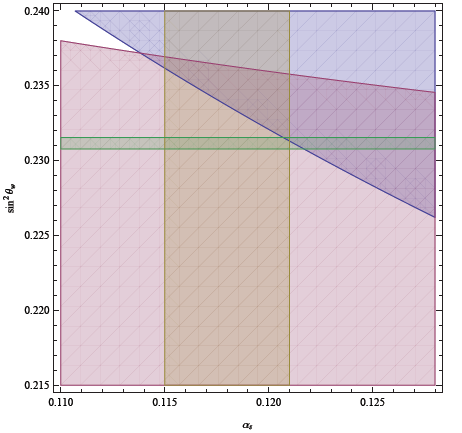} &
\includegraphics[scale=0.6]{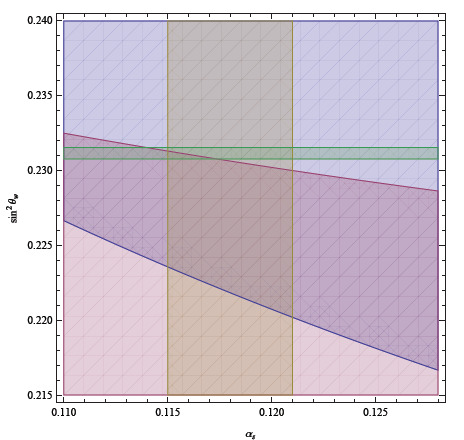} \\ 
Fig. 6 & Fig. 7 \\ 
\end{tabular}

\end{center}

\vspace{1cm}
\noindent
Same as for figures 2-4 for the more restrictive lower limits for
$M_X$ and $M_{B-L}$.

\newpage

\begin{center}

\begin{tabular}{c}
\includegraphics[scale=0.6]{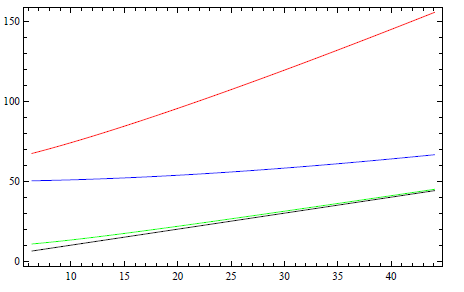} \\ 
 Fig. 8 \\ 
\end{tabular}

\vspace{0.2cm}
\noindent
Values of $m_1$ (black), $m_2$ (green), $m_3$ (blue) and $\Sigma m_i$ (red) vs. 
$|m_1|$ in meV.
\vspace{1cm}

\begin{tabular}{c}
\includegraphics[scale=0.6]{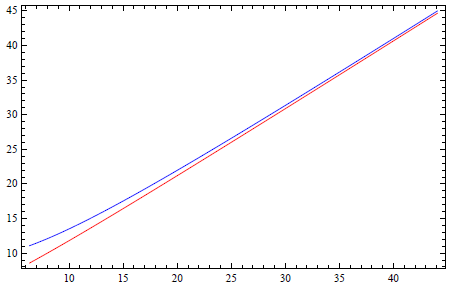} \\ 
 Fig. 9 \\ 
\end{tabular}

\vspace{0.2cm}
\noindent
Values of $m_{\nu_e}$ (blue) and $|m_{ee}|$ (red) vs. $|m_1|$ in meV.

\vspace{1cm}

\begin{tabular}{c}
\includegraphics[scale=0.6]{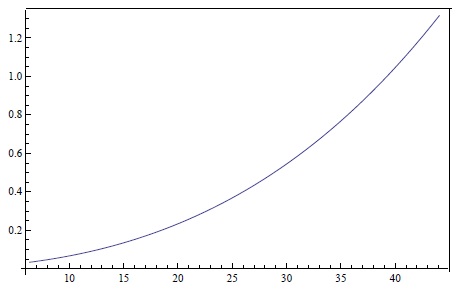} \\ 
 Fig. 10 \\ 
\end{tabular}

\vspace{0.2cm}
\noindent
Value of $\det(m_L)$ ($10^{-4}$ eV$^3$) in the allowed region for $|m_1|$ (meV).

\end{center}

\end{document}